\documentclass[12pt]{article}
\pdfoutput=1
\usepackage{amsmath, amssymb,amsthm,amscd,graphicx,dsfont,amsthm,youngtab}
\usepackage[table]{xcolor}
\usepackage{mathrsfs}
\usepackage{hyperref}
\usepackage{dynkin-diagrams}
\usepackage{longtable}
\usepackage{mathscinet}

\usepackage{tikz}
\usepackage{tikz-cd}
\usetikzlibrary{bending}

\pgfkeys{/Dynkin diagram, edge length=0.76 cm, root
    radius=.1cm}

\numberwithin{equation}{section}
\theoremstyle{definition}
\newtheorem{thm}{Theorem}[section]

\theoremstyle{definition}
\newtheorem{defn}{Definition}[section]

\theoremstyle{definition}
\newtheorem*{conj*}{Conjecture}

\theoremstyle{remark}

\newcommand{\cC}{\mathcal{C}}

\newcommand{\cN}{\mathcal{N}}
\newcommand{\cO}{\mathcal{O}}

\newcommand{\Z}{\mathbb{Z}}

\newcommand{\fh}{\mathfrak{h}}
\newcommand{\g}{\mathfrak{g}}
\newcommand{\fl}{\mathfrak{l}}

\newcommand{\lambdabar}{\overline{\lambda}}
\newcommand{\sigmabar}{\overline{\sigma}}

\DeclareMathOperator{\Vol}{Vol}
\DeclareMathOperator{\Tr}{Tr}
\DeclareMathOperator{\End}{\mathrm{End}}

\DeclareMathOperator{\Lie}{Lie}
\DeclareMathOperator{\ad}{ad}
\DeclareMathOperator{\Ad}{Ad}
\DeclareMathOperator{\Aut}{Aut}

\newcommand{\Zg}{\mathcal{Z}^G}
\newcommand{\dO}{\dim \cO}
\def\LG{{^L\negthinspace G}}

\begin{document}

\begin{titlepage}
\begin{flushright}
KIAS-P19058\\
\end{flushright}

\vskip 2cm

\begin{center}
{\large \bfseries
Supersymmetric Yang--Mills Matrix Integrals Revisited
}

\vskip 1.2cm
Richard Eager \footnote{reager@kias.re.kr}
\bigskip
\bigskip

School of Physics, Korea Institute for Advanced Study, Seoul 02455, Korea

\vskip 1.5cm

\textbf{Abstract}
\end{center}
We evaluate the twisted partition function of four-dimensional $\cN = 1$ supersymmetric Yang--Mills theory reduced to a point for all simple gauge groups.
The partition function is expressed as a sum of residues.  The types of residues that appear are classified by distinguished nilpotent orbits.  Surprisingly, the multiplicity of residues of each type is proportional to their common value.  The sum over residues has the same form as the Plancherel measure for Yang's system of particles.  Intriguingly, the precise constants appearing in the Plancherel measure also appear in the formal degrees of unipotent discrete series representations of $p$-adic groups.
\medskip
\noindent

\bigskip
\vfill
\end{titlepage}


\newpage 
\section{Introduction}
Supersymmetric quantum field theory has served as an import proving ground for testing new techniques that elucidate the dynamics of general quantum field theories.  A powerful tool for studying the dynamics of supersymmetric quantum field theories is supersymmetric localization.
Supersymmetric localization reduces infinite-dimensional path-integrals to ordinary finite-dimensional integrals.  For partition functions on flat space, supersymmetric localization can further reduce the partition function to a sum of residues.

The choice of integration contour is typically the most subtle aspect of supersymmetric localization.  Recently, a systematic procedure to derive
the contour for two-dimensional gauge theories was derived from localization of the path-integral \cite{Benini:2013nda, Benini:2013xpa}, expressing the supersymmetric partition function as a sum of Jeffrey--Kirwan residues.  This derivation was adapted to quantum mechanics in \cite{Hwang:2014uwa, Hori:2014tda}.

In this paper, we will evaluate the partition function of four-dimensional $\cN = 1$ supersymmetric Yang--Mills theory reduced to a point.
These matrix integrals have been extensively studied for over two decades, and many partial results have been obtained  \cite{Yi:1997eg, Sethi:1997pa, Green:1997tn, Krauth:1998xh, Moore:1998et, Kac:1999av, Pestun:2002rr, Lee:2016dbm, Lee:2017lfw, Hwang:2017nop}.  We find a simple formula that extends these previous results to all simple gauge groups.

As the rank of the gauge group increases, the number and complexity of the residues present a combinatorial challenge. 
In this paper, we utilize the theory of nilpotent orbits to classify the combinatorial types of the residues that appear in the localization calculation.  The residues naturally correspond to distinguished nilpotent orbits.  We make an intriguing conjecture about their multiplicity, which suggest the existence of a natural contour of integration that is simpler than the one obtained in \cite{Benini:2013xpa}.

One of the original motivations for studying supersymmetric matrix integrals comes from attempts to compute the Witten index of supersymmetric matrix quantum mechanics.  M-theory predicts a unique bound state for  $SU(N)$ $\cN = 16$ supersymmetric quantum mechanics \cite{Witten:1995ex, Yi:1997eg, Sethi:1997pa}.  This prediction can be tested by reducing the supersymmetric partition function --- the Witten index ---  to a sum of matrix model integrals \cite{Hwang:2017nop}.  The M-theory prediction has been generalized to other classical gauge groups in \cite{Hanany:1999jy}.
The Witten index and the matrix model partition functions have been intensely studied by supersymmetric localization \cite{Moore:1998et, Pestun:2002rr, Lee:2016dbm, Lee:2017lfw, Hwang:2017nop}, heat kernel methods \cite{Yi:1997eg, Green:1997tn, Kac:1999av}, the mass deformation method \cite{Porrati:1997ej, Kac:1999av}, and Monte Carlo simulation \cite{Krauth:1998xh}.

While the partition functions computed by the various different methods are equal, the intermediate expressions are surprisingly different.  In supersymmetric localization, the final result for the Witten index is related to a sum over nilpotent (or unipotent) orbits.  In the heat kernel method, the final result is given as a sum over elliptic Weyl group elements \cite{Kac:1999av}.  As a result, we find a mathematical identity relating nilpotent orbits to Weyl group representations.  Similar relations have previously been studied in the part of geometric representation theory known as Springer theory.

Remarkably, the same types of integrals play a prominent role in the theory of $p$-adic Lie groups and the Langlands program.  
The same residue calculations have previously appeared in determining the Plancherel measure on Yang's system of particles \cite{MR1432038, 2015arXiv151208566D}.
This integrable system is one of the original examples of the BPS/CFT correspondence \cite{Moore:1997dj, Gerasimov:2006zt, Gerasimov:2007ap} where it was connected to a two-dimensional version of supersymmetric Yang--Mills theory with four supercharges.
The Plancherel measure also has a natural $p$-adic version that \cite{MR1748271} we expect to be related to the equivariant Witten index of supersymmetric quantum mechanics \cite{Lee:2016dbm}.

In the next section, we will review the reduction of the matrix model partition function to a sum of residues by localization.
We then describe our formula for the partition function as a sum over distinguished nilpotent orbits.
In section \ref{sec:rootsystems}, we recall some of the terminology for root systems.
In section \ref{sec:nilpotentorbits}, we define semisimple and nilpotent orbits and explain how the classification of semi-simple orbits is used to classify nilpotent orbits.  We then make use of the theory of nilpotent orbits to classify the poles that appear in the localization calculation.  We list all of the relevant nilpotent orbit data used in the evaluation of the matrix model partition in Tables \ref{tab:A4}--\ref{tab:E8}.

\section{Localization Computation}
\label{sec:localization}
The aim of this paper is to evaluate the matrix integral:
\begin{equation}
\Zg = \frac{1}{\Vol(G/Z(G))} \int d\lambda \, \mathrm{d} A \, dD \; e^{-S_{YM}}
\end{equation}
where 
\begin{equation}
S_{YM} = - \Tr \left( \frac{1}{4} \left[ A_{\mu}, A_{\nu} \right]^2 + \lambdabar \sigmabar^{\mu} \left[ A_{\mu}, \lambda \right] - 2D^2 \right)
\end{equation}
is the action of four-dimensional $\cN = 1$ supersymmetric Yang--Mills theory with gauge group $G$ reduced to a point.
The normalization is by a factor $\Vol(G/Z(G))$ where $Z(G)$ is the center of $G$.
Here, $\lambda, A_{\mu},$ and $D$ are the reductions of the component fields in the four-dimensional $\cN = 1$ vector multiplet.
The original four-dimensional fields are a two complex Weyl spinors, a four-component vector, and a scalar.  Upon dimensional reduction, the fields of the vector become the four scalars $A_{\mu}.$  The fields are all valued in the adjoint representation of the Lie algebra of $G$.
We follow the normalization conventions in \cite{Pestun:2002rr}.

Supersymmetric localization reduces the problem to an integral over a Lagrangian submanifold $\cC$ of a Cartan subalgebra $\fh$ of $\g$:
\begin{equation}
\label{eq:Zdet}
\Zg = \frac{1}{|W_G|} \frac{1}{\Vol(T/Z(G))} \int_{\cC} d^{r} u  \frac{\det'(\ad(u))}{\det(\ad(u) + \epsilon)}
\end{equation}
where $r$ is the rank of $G$, $d^r u$ is the volume form on $\fh,$ $\ad(u)$ is the adjoint action of $\g,$ and $|W_G|$ is the order of the Weyl group of $G.$
Here $\det'(\ad(u))$ is the restriction of $\det(\ad(u))$ to $\g \setminus \fh$ \cite{Moore:1998et, Pestun:2002rr}.
The normalization is by a factor $\Vol(T/Z(G))$ where $T$ is the maximal torus of $G$.
The integrand arises as a ratio of one-loop determinants.  The parameter $\epsilon$ can either be inserted as an ad-hoc regulator \cite{Moore:1998et}, or be obtained as a limit of the $S^1$-equivariant parameter in supersymmetric quantum mechanics on reduction to zero dimensions \cite{Hwang:2017nop}. 

There are two proposals for the contour of integration.
The original contour proposal selects poles with positive imaginary part \cite{Moore:1998et, Pestun:2002rr}.
More recently, a different contour of integration was derived in  \cite{Benini:2013nda, Benini:2013xpa}.
The precise form of the contour depends on the choice of a Jeffrey--Kirwan covector.
However, in this particular problem, the result is independent of the choice of covector.
Once a contour is specified, the integral can be evaluated by summing the residues of the poles that are enclosed by the contour.

To determine the residues, consider the operator $ad(u) + \epsilon$ acting on an element $\phi$ of  $\g.$
Poles potentially occur when
\begin{equation}
\label{eq:nilp}
[\phi, u] = \epsilon \phi.
\end{equation}
Thus, $\phi$ and $u$ satisfy one of the relations of the $\mathfrak{sl}_2$ Lie algebra.  The relation~\eqref{eq:nilp} implies that
$\phi$ is nilpotent since $\g$ is finite dimensional.
A nilpotent element of $\mathfrak{su}_n$ can be described by the sizes of the blocks in its Jordan block decomposition.
However, only elements with a single Jordan block can have a non-zero residue \cite{Moore:1998et}.
For other groups, the classification of residues follows from the theory of nilpotent orbits.
A textbook account of the theory of nilpotent orbits is given in \cite{MR1251060}.
We will summarize and apply it to the classification of residues in section \ref{sec:nilpotentorbits}.

The residues can be evaluated by expanding $ad(u) + \epsilon$,
\begin{equation}
\Zg = \frac{1}{(2 \pi i \epsilon)^r} \frac{ \det || \alpha^{s}||}{|W_G|} \int_{\cC} d^{r} u  \prod_{\alpha} \frac{\alpha u}{\alpha u + \epsilon}
\end{equation}
where $\alpha^s$ is a basis of simple roots and the product is taken over all roots $\alpha$ of $\g.$
The eigenvalues of $\ad(u)$ on $\g$ can be determined from the eigenvalues of $\ad(u)$ on the simple roots of $\g$.
This information is recorded in the weighted Dynkin diagram of the nilpotent orbit.
On each representative $\phi_{\cO}$ of a nilpotent orbit $\cO$, we consider the order of vanishing of the numerator and denominator.
We will see that the order of the pole is bounded by the rank of $\g,$ with equality if and only if $\cO$ is a {\it distinguished nilpotent orbit}.
The residue simplifies to
$|A(\cO)|^{-1} \widehat{Res}(\cO)$
where
\begin{equation}
\widehat{Res}(\cO) = \frac{1}{\epsilon^r} \frac{\prod'_{\alpha} \alpha u }{\prod'_{\alpha} \alpha u + \epsilon},
\end{equation}
and the symbol $\prod'$ denotes the product of the non-zero terms.
$A(\cO)$ is the  {\it component group} of the stabilizer of the orbit, and $|A(\cO)|$ is its order.
Rather surprisingly, the multiplicity of the residue is
\begin{equation}
\label{eq:mult}
|W_G|  \widehat{Res}(\cO),
\end{equation}
where $|W_G|$ is the order of the Weyl group of $G$.  
Our main result is the following conjectural form of the supersymmetric Yang--Mills matrix model integral:  
\begin{equation}
\label{eq:Zg}
\Zg = \sum_{\cO} \frac{1}{|A(\cO)|} \widehat{Res}(\cO)^2,
\end{equation}
where the sum is over distinguished nilpotent orbits $\cO.$
The formula is obtained from summing the residues with their corresponding multiplicity.
The principle that {\it fixed points with extra $U(1)$ factors left unbroken do not contribute to $\Zg$} \cite{Moore:1998et} gives another heuristic reason for why only {\it distinguished nilpotent orbits} contribute to the sum.
Furthermore, a fixed point with an unbroken discrete symmetry group
$A(\cO)$ will have its contribution weighted by a factor of $|A(\cO)|^{-1}.$
While this precise factor has checked against direct residue calculations, it remains an interesting challenge to give a rigorous derivation.

The remainder of this paper is devoted to explaining the ingredients appearing in equation~\eqref{eq:Zg} needed to compute $\Zg.$
Using the values of $\widehat{Res}(\cO)$ and $|A(\cO)|$ for the distinguished nilpotent orbits in Tables \ref{tab:A4}--\ref{tab:E8}, we evaluate the matrix integral $\Zg$ and list the result for the rank four classical and exceptional Lie groups in Table~\ref{tab:Zg}.
The results for $E_7$ and $E_8$ are new and the other results agree with previous calculations \cite{Pestun:2002rr, Hwang:2017nop}.  
\begin{table}[h]
\begin{center}
\caption{$\Zg$ for the rank four classical and exceptional Lie groups.}
\label{tab:Zg}
\begin{tabular}{|c|r|}
\hline
$G$ & $\Zg$ \\
\hline
\hline
$SU(5)$ &  $\frac{1}{25}$\\
\hline
$SO(9)$ & $\frac{613}{8192}$ \\
\hline
$Sp(4)$ & $\frac{1275}{16384}$ \\
\hline
$SO(8)$ &  $\frac{117}{2048}$   \\
\hline
\end{tabular}
\qquad
\begin{tabular}{|c|r|}
\hline
$G$ & $\Zg$ \\
\hline
\hline
$G_2$ &  $\frac{151}{864}$\\
\hline
$F_4$ & $\frac{493013}{3981312}$ \\
\hline
$E_6$ & $\frac{14317}{209952}$ \\
\hline
$E_7$ &  $\frac{24826523}{254803968}$   \\
\hline
$E_8$ & $\frac{6304867107459827}{37150418534400000}$ \\
\hline
\end{tabular}
\end{center}
\end{table}%

\section{Root Systems}
\label{sec:rootsystems}
We now establish some notation for root systems.
A {\it Cartan subalgebra} $\fh$ of $\g$ is a maximal abelian subalgebra of $\g$ consisting of semisimple elements.
Fixing a Cartan subalgebra, we have the root space decomposition
\begin{equation}
\g = \fh \oplus \bigoplus_{\alpha \in \Phi} \g_{\alpha}
\end{equation}
where $\Phi$ is a root system in $\fh^{*}.$
Root systems can be abstractly defined in terms of an arbitrary real vector space $V$.  For us, $V = \fh^{*}$.
Let $(\cdot, \cdot)$ be a symmetric bilinear form on $V$.  We choose an orthonormal basis $e_i$ of $V$ with $(e_i, e_j) = \delta_{ij}.$

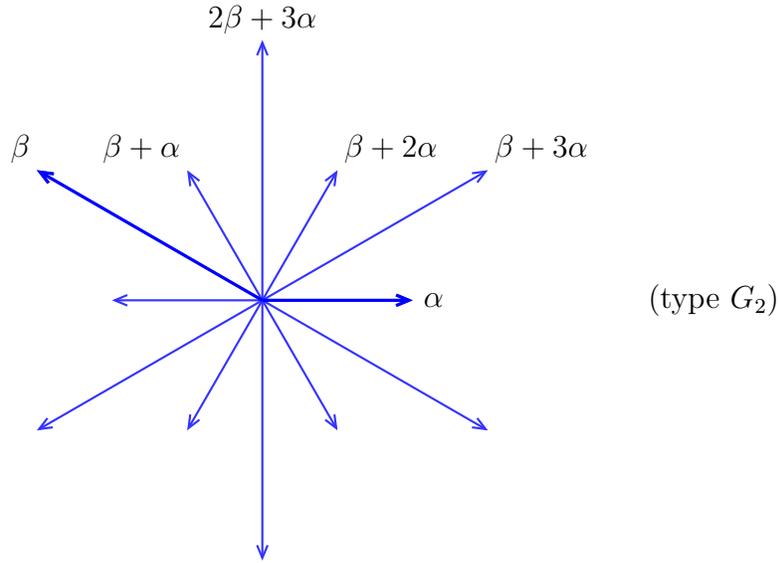
\begin{figure}
\def\LR{3.464}
\begin{center}
  \begin{tikzpicture}[
    -{Straight Barb[bend,
       width=\the\dimexpr10\pgflinewidth\relax,
       length=\the\dimexpr12\pgflinewidth\relax]},
  ]
    \foreach \i in {0, 1, ..., 5} {
      \draw[thick, blue!80] (0, 0) -- (\i*60:2);
      \draw[thick, blue!80] (0, 0) -- (30 + \i*60:\LR);
    }
      \draw[very thick, blue] (0, 0) -- (0*60:2);
      \draw[very thick, blue] (0, 0) --(5*30:\LR);
    \node[right] at (2, 0) {$\alpha$};
       \node[above right, inner sep=.2em] at (1*30:\LR) {$\beta + 3\alpha$};
       \node[above right, inner sep=.2em] at (2*30:2) {$\beta + 2\alpha$};
       \node[above, inner sep=.2em] at (3*30:\LR) {$2\beta + 3\alpha$};
      \node[above left, inner sep=.2em] at (4*30:2) {$\beta + \alpha$};
    \node[above left, inner sep=.2em] at (5*30:\LR) {$\beta$};
    \node at (6, 0) {(type $G_2$)};

  \end{tikzpicture}
      \caption{Root system for $G_2$.}
    \label{fig:G2}
  \end{center}
  \end{figure}

A {\it positive simple root system} is a subset $\Delta$ of a root system $\Phi$ that satisfies the following conditions:
\begin{itemize}
\item $\Delta$ is a basis for the vector space $V$,
\item Each element $\beta \in \Phi$ can be 
\begin{equation}
\label{eq:linear}
\beta = \sum_{\alpha \in \Delta} m_{\alpha} \alpha,
\end{equation}
where the $m_{\alpha}$ are either all non-negative integers or all non-positive integers.
\end{itemize}
The elements of $\Delta$ are called {\it positive simple roots}.
From $\Delta,$ we can form the {\it Dynkin diagram} of $\g.$  The vertices are given by the elements of $\Delta,$ and the number of edges between two vertices is determined by the angle between the two corresponding elements of $\Delta.$ 
The root systems for $B_n, C_n, $ and $D_n$ are \cite{MR1251060}
\begin{align}
B_n & = \left\{ \pm e_i \pm e_j \vert 1 \le i< j \le n \right\} \cup \left\{\pm e_i \vert 1 \le i \le n \right\} \\
C_n & = \left\{ \pm e_i \pm e_j \vert 1 \le i< j \le n \right\} \cup \left\{\pm 2 e_i \vert 1 \le i \le n \right\} \\
D_n & = \left\{ \pm e_i \pm e_j \vert 1 \le i< j \le n \right\}.
\end{align}
The root system for $G_2$ is shown in Figure~\ref{fig:G2}.
The positive simple roots for $B_n, C_n,$ and  $D_n$ are:
\begin{align}
B_n & = \left\{ e_1 - e_2, e_2 - e_3, \dots, e_{n-1} - e_n, e_n \right\} \\
C_n & = \left\{ e_1 - e_2, e_2 - e_3, \dots, e_{n-1} - e_n, 2e_n\right\} \\
D_n & = \left\{ e_1 - e_2, e_2 - e_3, \dots, e_{n-1} - e_n, e_{n-1} + e_n\right\}.
\end{align}
The positive simple roots for $G_2$ are $\alpha$ and $\beta$.
\section{Nilpotent Orbits}
\label{sec:nilpotentorbits}
In this section, we define semisimple and nilpotent elements in a Lie algebra and their corresponding orbits.
A linear operator $X$ acting on a finite-dimensional vector space $V$ can be viewed as an element of $\End(V).$
It is a {\it semisimple} operator if every $X$-invariant subspace has an $X$-invariant complement.
It is a {\it nilpotent} operator if $X^r \equiv \underbrace{X \circ \dots \circ X}_{r} = 0.$
We adapt these notions to the case where $V$ is a complex semisimple Lie algebra $\g.$
In a complex semisimple Lie group $G$, the {\it adjoint representation}
is defined by the action
\begin{align}
\ad: \g & \mapsto \End(\g) \\
\label{eq:adjointaction}
\ad_X Y & = [X,Y].
\end{align}
An element $X \in \g$ is {\it nilpotent} if $\ad_X$ is a nilpotent endomorphism of the vector space $\g.$  Similarly, $X \in \g$ is {\it semisimple} if $ad_X$ is semisimple.
These notions let us generalize the Jordan decomposition theorem to arbitrary semisimple groups.
\begin{thm}[{Jordan Decomposition \cite[theorem 1.1.1]{MR1251060}}]
If $\g$ is semisimple, every element $X \in \g$ can be written uniquely in the form
$X = X_s + X_n$, with $X_s$ is a semisimple element, $X_n$ is a nilpotent element, and $[X_s, X_n] = 0.$
\end{thm}

We can now define semisimple and nilpotent orbits under the adjoint action of $G.$
Recall that the adjoint representation of $G,$ 
\begin{equation}
Ad : G \rightarrow \Aut(\g)
\end{equation}
is defined by the derivative of the conjugation automorphism $\Psi_g : G \rightarrow G,$ where  $\Psi_x(y) = x y x^{-1}.$
\begin{defn}[{\cite[pp. 9]{MR1251060}}]
The {\it orbit though} $X$ is the set of elements
$$\cO_X \equiv G_{ad} \cdot X = \left\{\phi \cdot X, \phi \in G_{ad} \right\}$$
where $G_{ad}$ is the identity component of $\Aut(\g).$
\end{defn}
The orbit $\cO_X$ is called a {\it nilpotent orbit} if $X$ is a nilpotent element of $\g$. Similarly, an orbit $\cO_X$ is called a {\it semisimple orbit} 
if $X$ is semisimple.  These definitions are independent of the choice of element $X$ labeling the orbit.

\subsection{Semisimple Orbits}
The structure of semisimple orbits is much simpler than that of nilpotent orbits.  The general strategy to classify nilpotent orbits is to relate them to semisimple orbits.
\begin{thm}[{\cite[theorem 2.2.4]{MR1251060}}]
Let $\g$ be a semi-simple Lie algebra with Cartan $\fh$ and Weyl group $W$.  Then there is a bijective correspondence:
\[
\left\{\text{semi-simple orbits} \right\}  \longrightarrow \fh/W.
\]
\end{thm}
We can therefore parameterize semi-simple orbits by a fundamental domain for $\fh/W$. One choice is given by the following:
\begin{defn}[{\cite[theorem 2.2.7]{MR1251060}}]
A fundamental domain for $\fh/W$ is:
\[
D_{\Delta} = \left\{ x \in \fh \; \Big\vert 
\!\begin{aligned}
       & \; \Re(\alpha(x)) \ge 0 \text{ and whenever } \Re(\alpha(x)) = 0, \\ 
       & \text{ then } \Im(\alpha(x)) \ge 0 \text{ for all } \alpha \in \Delta
    \end{aligned} 
 \right\}.
\]
\end{defn}

\subsection{Classification of Nilpotent Orbits}
To each nilpotent element we can associate a {\it standard triple}.  It will allow us to use the previously discussed structure theory of semisimple orbits to classify nilpotent orbits.
\begin{defn}
\label{def:triple}
A {\it standard triple} $\left\{e,h,f\right\}$ is a basis of a $\mathfrak{sl}_2$-subalgebra of $\g$ satisfying the standard commutation relations:
$$[h, e] = 2e, \; [h,f] = -2f, \; [e,f] = h.$$
\end{defn}
A fundamental result for the structure of nilpotent orbits is given by the following theorem.
\begin{thm}[Jacobson--Morozov]
Let $\g$ be a complex semisimple Lie group.  Then any nilpotent element $X \in \g$ can be extended to a standard triple $\left\{e,h,f\right\}$.
\end{thm}
From this theorem it follows that there is a bijective correspondence:
\begin{align*}
\left\{\text{Conjugacy classes of } \mathfrak{sl}_2\text{-triples} \right\} &  \longrightarrow \left\{\text{Nilpotent orbits } \cO_e \right\} \\
(e,h,f) &  \longrightarrow e.
\end{align*}
Given a nilpotent element $e$ and corresponding standard triple $(e,h,f),$ we can decompose $\g$ into its $ad_h$ eigenspaces.
\[
\g = \bigoplus_{i \in \Z} \g_i,
\]
where the sum is over the integral eigenvalues of $ad_h$.  Integrality follows from the representation theory of $\mathfrak{sl}_2$.

We now apply the above structure theory to determine the poles in Equation \eqref{eq:Zdet} that contribute to the matrix integral $\Zg.$
Recall that the potential poles are described by elements $\phi$ and $u$ satisfying Equation \eqref{eq:nilp} that are part of
a standard triple with $\phi$ nilpotent.
The number of zeroes in the numerator is $\dim(\g_0) - \dim(\fh),$
and the number of zeroes in the denominator is $\dim(\g_2).$ 
\footnote{There is a factor of 2 relating the $ad_h$ eigenvalues and the coefficient of $\epsilon$ used in localization.}

\begin{thm}[{\cite[theorem 8.2.1]{MR1251060}}]
 $\dim(\g_0) \ge \dim(\g_2)$ with equality if and only if $e$ is a {\it distinguished} nilpotent element.
\end{thm}

Therefore the poles in the integrand with non-vanishing residues correspond to distinguished nilpotent orbits.
While for our purposes we can take $\dim(\g_0) = \dim(\g_2)$ to be the definition of a nilpotent orbit, the standard definition is:
\begin{defn}
A nilpotent orbit $\cO_e$ is {\it distinguished} if the smallest Levi subalgebra containing $e$ is $\g$ itself.
\end{defn}
\begin{thm}[Bala--Carter]
There is a bijection
$$
\left\{\text{Nilpotent orbits } \cO_e \right\} \longleftrightarrow \left\{ \text{Pairs } (\fl, \cO_e^{\fl} ) \right\} 
$$
between nilpotent orbits $\cO_e$ and pairs consisting of a Levi subalgebra $\fl$ of $\g$ and a distinguished nilpotent orbit $\cO_e^{\fl}$ of $\fl.$
\end{thm}
The Bala--Carter theorem reduces the classification of nilpotent orbits to the classification of distinguished nilpotent orbits.

\subsection{Weighted Dynkin Diagrams}
A convenient way to label a nilpotent orbit is by its weighted Dynkin diagram.
For each nilpotent orbit $\cO_e$ there is a corresponding semi-simple orbit $\cO_h$ where $h$ is the element of the $\mathfrak{sl}_2$-triple containing $e.$
Fixing a set basis positive simple roots $\Delta,$ we can choose a conjugate $\mathfrak{sl}_2$-triple such that the corresponding $h$ lives in the fundamental domain $D_{\Delta}.$
The $h$ in the fundamental domain satisfies:
\begin{equation}
\alpha(h) \in \left\{0,1,2 \right\} \text{ for all } \alpha \in \Delta.
\end{equation}
The {\it weighted Dynkin diagram} of a nilpotent orbit $\cO_e$ is obtained by listing the eigenvalues
$\alpha(h)$ for each vertex $\alpha$ of the Dynkin diagram.
Distinguished nilpotent orbits have $\alpha(h) \in  \left\{0, 2 \right\}$ for all $\alpha \in \Delta$.
We can therefore compute the eigenvalues of $\ad_h$ on $\g$ using the decomposition in Equation \eqref{eq:linear}.
Therefore, we can compute $\widehat{Res}(\cO)$ from its weighted Dynkin diagram.

We now give an example of the computation of $\widehat{Res}(\cO)$.  For each Lie algebra $\g,$ there is always a unique {\it principal nilpotent orbit} $\cO_{prin},$ that is always distinguished and of maximal dimension.  For this orbit, the weighted Dynkin diagram has $\alpha(h) = 2$ for all $\alpha \in \Delta$.
The action of $\ad_h$ on $\g$ decomposes into representations of $\mathfrak{sl}_2$ of dimension $2e_i - 1$ where $e_i$ are the degrees of the fundamental invariants of $\g$ {\cite[theorem 4.4.11]{MR1251060}}.  Therefore,  $\widehat{Res}(\cO)$ can be evaluated explicitly as
\begin{equation}
\widehat{Res}(\cO_{prin}) = \prod_{i} \frac{e_i - 1}{e_i}.
\end{equation}
The factors in the numerator and denominator have interesting interpretations as well.  The product of the degrees of the fundamental invariants equals the order of the Weyl group $|W_G|$. The numerator equals the number of elliptic Weyl group elements $|W^{ell}_G|$. 
The final result is
\begin{equation}
\widehat{Res}(\cO_{prin}) = \frac{|W^{ell}_G|}{|W_G|}.
\end{equation}
The multiplicity obtained from combining the previous equation and equation \eqref{eq:mult} is $|W^{ell}_G|.$
It is naturally interpreted as the top-dimensional cohomology of the Orlik--Solomon algebra of the Coxeter hyperplane arrangement corresponding to $G$ \cite{MR0422674, MR1217488}.
\section{Distinguished Nilpotent Orbit Data}
In this section, we describe the results of the classification of nilpotent orbits.  For classical Lie algebras, the distinguished nilpotent orbits
have a simple description in terms of partitions.
\begin{thm}[{\cite[theorem 8.2.14]{MR1251060}}]
For $A_n$, only the principal nilpotent orbit is distinguished.  In types $B_n, C_n,$ and $D_n,$ nilpotent orbits correspond to partitions with distinct odd, even, and odd parts of $2n+1,2n,$ and $2n,$ respectively.  
\end{thm}

From the partition, the weighted Dynkin diagram, $\widehat{Res}(\cO), A(\cO),$ and $\dO$ can easily be obtained from the results in \cite{MR1251060}.
We give examples for the classical Lie algebras of rank four.  We also lie a representative of the nilpotent element $e$ in the standard triple.  This is not needed to evaluate $\Zg$, but it is useful to compare to the direct computation of residues.  For the exceptional Lie algebras, standard triples are given in \cite{MR2434879}.
The unbroken gauge symmetry $A(\mathcal{O}_X)$ is obtained from {\cite[corollary 6.1.6]{MR1251060}}.  It is either $Z_2^n$ or a symmetric group $S_n$ on $n$ letters.
 \begin{thm}[{\cite[theorem 4.1.3]{MR1251060}}]
The dimension of a nilpotent orbit is 
\[
dim(\cO) = dim(\g) - dim(\g_0) - \dim(\g_1).
\]
\end{thm}

\begin{thm}[{\cite[theorem 8.2.3]{MR1251060}}]
If $e \in \g$ is a distinguished nilpotent element, then $\dim(\g_1) = 0$
and hence 
\[
dim(\cO_e) = dim(\g) - dim(\g_0). 
\]
\end{thm}
For each classical Lie group of rank four and all of the exceptional Lie groups, we list the corresponding nilpotent orbits along with their standard label and the data $\widehat{Res}(\cO), A(\cO), \dO$ in Tables \ref{tab:A4}--\ref{tab:E8}. 

\section{Conclusions and Further Speculations}
We have evaluated the supersymmetric matrix partition function $\Zg$ for all simple Lie groups $G.$
Our final result in Equation~\eqref{eq:Zg} suggests the existence of a natural contour of integration and connections to nilpotent and unipotent representations of $p$-adic groups.  We hope that the existence of such a simple form of the result and contour will find applications in other supersymmetric localization calculations and allow for new calculations where the gauge group has large rank.  In particular, it should help compute exact results in the limit of large rank.

There is a long history of partition functions that can be computed by either point counting over finite fields or by localization methods.
For example, the cohomology of moduli spaces of vector bundles has been computed using Deligne's proof of the Weil conjectures \cite{MR262246, MR364254, MR429897}.
It involves counting the number of points in the moduli space defined over a finite field.
A different calculation by Atiyah and Bott used the Yang--Mills functional on a Riemann surface as a perfect Morse function and localized the computation to the critical set of the Yang--Mills functional \cite{MR702806}.
The common theme is that the order of the finite field in the point-counting approach appears in the same form as the equivariant parameter in localization calculations.
In revisiting the work of Atiyah and Bott, Witten developed non-abelian localization \cite{Witten:1992xu}.
A mathematical formalization led to the notion of the Jeffrey--Kirwan residue \cite{JeffreyKirwan, SzenesVergne}, which plays a key role in evaluating the matrix model partition function $\Zg.$

The tantalizing connection to our problem is the following correspondence of Lusztig.
Given a simple group $G$ and its Langlands dual group $\LG,$ there is bijection between a certain class of irreducible unipotent representations of $G$ and the set of triples $(s,N, \rho)$ 
where $s \in \LG$ is semisimple, $N \in \Lie(G)$ is nilpotent with $\Ad(s) N = qN,$ and $\rho$ is an irreducible representation of the finite group
$Z_{\LG}(s,N)/Z^0_{\LG}(s,N)$ \cite{MR1369407}.
The condition $\Ad(s) N = qN$ is precisely the form of the poles in the equivariant supersymmetric quantum mechanics partition function.
We plan to give a new computation of the supersymmetric Witten index in terms of unipotent orbits in a sequel.
A previous calculation using heat kernel methods results in a sum over elliptic Weyl group elements \cite{Kac:1999av, Hwang:2017nop}.
Thus, we expect a new relation between unipotent representations and elliptic conjugacy classes in the Weyl group.
Let $\alpha_0 = \sum_i \delta_i \alpha_i$ be the highest root of $\g$.  Denote the affine Dynkin diagram of $G$ by $\widehat{\Gamma_G}$ and let $H^i$ be the product group corresponding to removing vertex $i$ from the affine Dynkin diagram.  Then the equality between the heat kernel and localization calculations results in the following identity:
\begin{equation}
\sum_{w \in W^{ell}_G} \frac{1}{\det(1 - w)} = \frac{1}{Z(G)} \sum_{i \in \widehat{\Gamma_G}} \frac{1}{\delta_i} \mathcal{Z}^{H_i}.
\end{equation}
Perhaps it will shed new light on a similar correspondence of Lusztig \cite{MR3251707} and lead to a new connection between number theory and physics.
\section*{Acknowledgments}
The author would like to thank James Humphreys, Chiung Hwang, Seung-Joo Lee, Eric Sommers, and Piljin Yi for valuable discussions and communications.

\newpage
\rowcolors{0}{}{gray!10}
\begin{table}
\caption{Distinguished nilpotent orbits in the Lie algebra of type $A_4$.} 
\begin{center}
\begin{tabular}{|l|c|c|ccc|}
\hline
Label 				& Diagram 		& Representative & $\widehat{Res}(\cO)$ & $A(\cO)$ & $\dO$\\
					&  \dynkin{A}{ooooo} 	& & & & \\
\hline

$[5]$ 	& 2~~~~2~~~~2~~~~2~~~~2 	& $e_1-e_2, e_2 - e_3, e_3 - e_4, e_4 - e_5$ & $\frac{1}{5}$ & $1$ & 20\\

\hline
\end{tabular}
\end{center}
\label{tab:A4}
\end{table}
\rowcolors{0}{}{gray!10}
\begin{table}[htp]
\caption{Distinguished nilpotent orbits in the Lie algebra of type $B_4$.}
\begin{center}
\begin{tabular}{|l|c|c|ccc|}
\hline
Label 				& Diagram 		& Representative & $\widehat{Res}(\cO)$ & $A(\cO)$ & $\dO$\\
					&  \dynkin{B}{oo*o} 	& & & & \\
\hline

$[5, 3, 1]$ 	& 2~~~~0~~~~2~~~~0 	& $e_1 - e_2, e_2, e_3 - e_4, e_3 + e_4$ & $\frac{1}{64}$ & $Z_2^2$ & 28\\

$[9]$ 			& 2~~~~2~~~~2~~~~2	 &  $e_1 - e_2, e_2 - e_3, e_3 - e_4, e_4$ & $\frac{35}{128}$ & $1$ & 32 \\

\hline
\end{tabular}
\end{center}
\label{tab:B4}
\end{table}

\rowcolors{0}{}{gray!10}
\begin{table}[htp]
\caption{Distinguished nilpotent orbits in the Lie algebra of type $C_4$.}
\begin{center}
\begin{tabular}{|l|c|c|ccc|}
\hline
Label 				& Diagram 		& Representative & $\widehat{Res}(\cO)$ & $A(\cO)$ & $\dO$\\
					&  \dynkin{C}{ooo*} 	& & & & \\
\hline

$[6,2]$ 	& 2~~~~2~~~~0~~~~2 	& $e_1-e_2, e_2 - e_3, 2 e_3, 2e_4$ & $\frac{5}{64}$ & $Z_2$ & 30\\

$[8]$ 			& 2~~~~2~~~~2~~~~2	 & $e_1-e_2, e_2 - e_3, e_3 - e_4, 2e_4$ & $\frac{35}{128}$ & $1$ & 32 \\

\hline
\end{tabular}
\end{center}
\label{tab:C4}
\end{table}

\rowcolors{0}{}{gray!10}
\begin{table}[htp]
\caption{Distinguished nilpotent orbits in the Lie algebra of type $D_4$.}
\begin{center}
\begin{tabular}{|l|c|c|ccc|}
\hline
Label 				& Diagram 		& Representative & $\widehat{Res}(\cO)$ & $A(\cO)$ & $\dO$\\
					&  \dynkin{D}{oooo} 	& & & & \\
\hline

$[5,3]$ 			& 2~~~~0~~~~2~~~~2 	& $e_1 - e_2, e_1 - e_3, e_3 - e_4, e_3 + e_4$ & $\frac{3}{64}$ & $1$ & 22\\

$[7,1]$ 			& 2~~~~2~~~~2~~~~2	 & $e_1 - e_2, e_2 - e_3, e_3 - e_4, e_3 + e_4$ & $\frac{15}{64}$ & $1$ & 24 \\

\hline
\end{tabular}
\end{center}
\label{tab:D4}
\end{table}
\rowcolors{0}{}{gray!10}
\begin{table}[htp]
\caption{Distinguished nilpotent orbits in the Lie algebra of type $G_2$.}
\begin{center}
\begin{tabular}{|l|c|ccc|}
\hline
Label						& Diagram 				& $\widehat{Res}(\cO)$  	& $A(\cO)$ 		& $\dO$		\\
					 	& \dynkin{G}{*o}             		&					& 		&			 \\
\hline
$G_2(a_1)$			 	& 2~~~~0 				 	& $\frac{1}{12}$					&$S_3$ 	& 12 \\
$G_2$					& 2~~~~2 				 	& $\frac{5}{12}$					& $1$ 	& 14 \\
\hline
\end{tabular}
\end{center}
\label{tab:G2}
\end{table}
\rowcolors{0}{}{gray!10}
\begin{table}[htp]
\caption{Distinguished nilpotent orbits in the Lie algebra of type $F_4$.}
\begin{center}
\begin{tabular}{|l|c|ccc|}
\hline
Label 				& Diagram 		& $\widehat{Res}(\cO)$ & $A(\cO)$ & $\dO$\\
					&  \dynkin{F}{o*oo} 	 & & & \\
\hline

$F_4(a_3)$		 	& 0~~~~2~~~~0~~~~0 	 & $\frac{1}{576}$ & $S_4$ & 40\\

$F_4(a_2)$ 			& 0~~~~2~~~~0~~~~2	  & $\frac{5}{144}$ & $S_2$ & 44 \\

$F_4(a_1)$			& 2~~~~2~~~~0~~~~2 	 & $\frac{175}{1152}$ & $S_2$ & 46 \\

$F_4$				& 2~~~~2~~~~2~~~~2	   & $\frac{385}{1152}$ &$1$ & 48\\

\hline
\end{tabular}
\end{center}
\label{tab:F4}
\end{table}
\rowcolors{0}{}{gray!10}
\begin{table}[htp]
\caption{Distinguished nilpotent orbits in the Lie algebra of type $E_6$.}
\begin{center}
\begin{tabular}{|l|c|ccc|}
\hline
Label & Diagram					 &  $\widehat{Res}(\cO)$ & $A(\cO)$ & $\dO$ \\
	&  \dynkin{E}{oooooo}          	 &  & & \\
\hline

$E_6(a_3)$ 	& $2~~~~0~~~~\overset{\text{\normalsize 0}}{2}~~~~0~~~~2$	 & $\frac{1}{108}$ & $S_2$ & 66 \\

$E_6(a_1)$ 	& $2~~~~2~~~~\overset{\text{\normalsize 2}}{0}~~~~2~~~~2$ 	& $\frac{35}{324}$ & $1$ & 70 \\

$E_6$ & 		$2~~~~2~~~~\overset{\text{\normalsize 2}}{2}~~~~2~~~~2$	  & $\frac{77}{324}$ & $1$ & 72 \\

\hline
\end{tabular}
\end{center}
\label{tab:E6}
\end{table}
\rowcolors{0}{}{gray!10}
\begin{table}[htp]
\caption{Distinguished nilpotent orbits in the Lie algebra of type $E_7$.}
\begin{center}
\begin{tabular}{|l|c|ccc|}
\hline
Label & Diagram 				&  $\widehat{Res}(\cO)$ & $A(\cO)$ & $\dO$\\
	&  \dynkin{E}{ooooooo}           	& 			 & & \\
\hline

$E_7(a_5)$ & $0~~~~0~~~~\overset{\text{\normalsize 0}}{2}~~~~0~~~~0~~~~2$ &   $\frac{1}{4608}$ & $S_3$ & 112 \\

$E_7(a_4)$ & $2~~~~0~~~~\overset{\text{\normalsize 0}}{2}~~~~0~~~~0~~~~2$ &  $\frac{25}{9216}$ & $S_2$ & 116 \\

$E_7(a_3)$ & $2~~~~0~~~~\overset{\text{\normalsize 0}}{2}~~~~0~~~~2~~~~2$ &  $\frac{245}{9216}$  & $S_2$ & 120 \\

$E_7(a_2)$ & $2~~~~2~~~~\overset{\text{\normalsize 2}}{0}~~~~2~~~~0~~~~2$ &  $\frac{539}{9216}$ & $1$ & 122  \\

$E_7(a_1)$ & $2~~~~2~~~~\overset{\text{\normalsize 2}}{0}~~~~2~~~~2~~~~2$ & $\frac{715}{4608}$ & $1$ &  124 \\

$E_7$ & $2~~~~2~~~~\overset{\text{\normalsize 2}}{2}~~~~2~~~~2~~~~2$ &  $\frac{2431}{9216}$  & $1$  & 126 \\

\hline
\end{tabular}
\end{center}
\label{tab:E7}
\end{table}
\rowcolors{0}{}{gray!10}
\begin{table}[htp]
\caption{Distinguished nilpotent orbits in the Lie algebra of type $E_8$.}
\begin{center}
\begin{tabular}{|l|c|ccc|}
\hline
Label & Diagram 					&  $\widehat{Res}(\cO)$ & $A(\cO)$ & $\dO$\\
	&  \dynkin{E}{oooooooo}                 	&				&	&  \\
\hline

$E_8(a_7)$ & $0~~~~0~~~~\overset{\text{\normalsize 0}}{0}~~~~2~~~~0~~~~0~~~~0$ & 
$\frac{1}{8294400}$ & $S_5$ & 208 \\

$E_8(b_6)$ & $0~~~~0~~~~\overset{\text{\normalsize 0}}{2}~~~~0~~~~0~~~~0~~~~2$ & 
$\frac{875}{3981312}$ &$S_3$ & 220 \\

$E_8(a_6)$ & $0~~~~0~~~~\overset{\text{\normalsize 0}}{2}~~~~0~~~~0~~~~2~~~~0$ & 
$\frac{729}{409600}$ &$S_3$ & 224 \\

$E_8(b_5)$ & $0~~~~0~~~~\overset{\text{\normalsize 0}}{2}~~~~0~~~~0~~~~2~~~~2$ & 
$\frac{3773}{1555200}$ &$S_3$ & 226 \\

$E_8(a_5)$ & $2~~~~0~~~~\overset{\text{\normalsize 0}}{2}~~~~0~~~~0~~~~2~~~~0$ & 
$\frac{9625}{1327104}$ &$S_2$ & 228 \\

$E_8(b_4)$ & $2~~~~0~~~~\overset{\text{\normalsize 0}}{2}~~~~0~~~~0~~~~2~~~~2$ & 
$\frac{17875}{995328}$ &$S_2$ & 230 \\

$E_8(a_4)$ & $2~~~~0~~~~\overset{\text{\normalsize 0}}{2}~~~~0~~~~2~~~~0~~~~2$ & 
$\frac{539539}{11059200}$ &$S_2$ & 232\\

$E_8(a_3)$ & $2~~~~0~~~~\overset{\text{\normalsize 0}}{2}~~~~0~~~~2~~~~2~~~~2$ & 
$\frac{31603}{518400}$ &$S_2$ & 234 \\

$E_8(a_2)$ & $2~~~~2~~~~\overset{\text{\normalsize 2}}{0}~~~~2~~~~0~~~~2~~~~2$ & 
$\frac{3556553}{24883200}$ &$1$ & 236\\

$E_8(a_1)$ & $2~~~~2~~~~\overset{\text{\normalsize 2}}{0}~~~~2~~~~2~~~~2~~~~2$ & 
$\frac{7436429}{33177600}$&$1$ & 238 \\

$E_8$ & $2~~~~2~~~~\overset{\text{\normalsize 2}}{2}~~~~2~~~~2~~~~2~~~~2$ & 
$\frac{30808063}{99532800}$ &$1$ & 240  \\
\hline
\end{tabular}
\end{center}
\label{tab:E8}
\end{table}

\cleardoublepage 
\bibliographystyle{ytphys}
\bibliography{MatrixIntegrals}
\end{document}